\documentclass[prb,twocolumn,showpacs,preprintnumbers,amsmath]{revtex4}
\usepackage{dcolumn}
\usepackage{bm}
\usepackage{graphicx}
\usepackage{color}
\begin{document}

\title{Flux jumps and vortex pinning in Ba$_{0.65}$Na$_{0.35}$Fe$_2$As$_2$ single crystals}

\author{A. K. Pramanik}\email{ashpramanik@gmail.com}\affiliation{Institute for Solid State Research, IFW Dresden, D-01171 Dresden, Germany}
\author{S. Aswartham}\affiliation{Institute for Solid State Research, IFW Dresden, D-01171 Dresden, Germany}
\author{A. U. B. Wolter}\affiliation{Institute for Solid State Research, IFW Dresden, D-01171 Dresden, Germany}
\author{S. Wurmehl}\affiliation{Institute for Solid State Research, IFW Dresden, D-01171 Dresden, Germany}
\author{V. Kataev}\affiliation{Institute for Solid State Research, IFW Dresden, D-01171 Dresden, Germany}
\author{B. B\"{u}chner}\affiliation{Institute for Solid State Research, IFW Dresden, D-01171 Dresden, Germany}

\begin{abstract}
In this work we present the results of the bulk magnetization measurements in a superconducting state of single crystals of Ba$_{0.65}$Na$_{0.35}$Fe$_2$As$_2$. The isothermal magnetic field ($H||c$ axis) dependent magnetization ($M$) loops exhibit a second peak (SP) or `fishtail effect', as well as remarkable flux jumps at low temperatures. The critical current density $J_c$ obtained from the $M(H)$ loops is rather high, of the order of 10$^6$ A/cm$^2$. The analysis of the temperature and field dependent $J_c$ implies that high $J_c$ is mainly due to collective (weak) pinning of vortices by dense microscopic point defects with some contribution from a strong pinning mechanism. Pronounced magnetic instabilities in terms of flux jumps depend strongly on temperature as well on the field sweep rate. The field for first flux jump as calculated from an adiabatic model, however, is much lower than the experimentally observed values, and this enhanced stability is attributed to a flux creep phenomenon. The analysis of field dependent magnetic relaxation data additionally supports a collective pinning model. The data further suggests that SP in M(H) is likely related to crossover in creep dynamics from elastic to plastic mechanism. We have constructed the vortex phase diagram on field-temperature plane.     
\end{abstract}


\maketitle
\section{Introduction}
The magnetic instability or the so-called `flux jumps' is an important factor for the practical applications of superconductors along with the high critical temperature ($T_c$), high critical current density ($J_c$) and high upper critical field ($H_{c2}$); though all these issues are also of great interest for fundamental research. In type-II superconductors, above the lower critical field $H_{c1}$, the magnetic field ($H$) penetrates the bulk of superconductor in the form of vortices or flux lines, while each vortex or flux line consists of normal core surrounded by a whirlpool of supercurrent.\cite{tinkham} In case of perfect superconductors, the intervortex interaction forces the vortices to arrange in regular triangular lattice. However, the presence of disorder acts as pinning centers that hinder the vortex motion, hence the interplay between these localizing and delocalizing forces results in a nonequilibrium situation creating a condition called `critical state'. Under certain circumstances like thermal fluctuations, the critical state becomes unstable where vortices rush into the volume of superconductor. This event leads to magnetic instability and shows up as a discontinuity in field dependent magnetization ($M$) measurements.\cite{wipf,altshuler} This process suddenly reduces $J_c$ appreciably and drives the superconductor towards normal state, which is unfavorable for technological applications. The flux jumps phenomenon has been observed and studied in great details in both conventional low temperature Nb based superconductors\cite{wipf,werth} as well as high temperature superconductors like cuprates,\cite{muller,henry, nabialek} MgB$_2$,\cite{victor,salazar} etc. 

The recent discovery\cite{kamihara} of superconductivity (SC) in Fe-based pnictides has ushered a hope for possible technological applications as these materials exhibit high $T_c$, type-II nature, large $H_{c2}$ and comparatively low anisotropy.\cite{johnston} The critical current density and vortex dynamics have extensively been studied for all families of Fe-pnictides. The studies show reasonably high $J_c$ in the range of 10$^{-5}$ to 10$^{-6}$ A/cm$^{2}$ at low temperature.\cite{beek-1111, ashim-LiFeAs,prozorov,yang,sun,yama,wang,kope,shen,ashim-CFCA,beek-prl,civale,taen} Interestingly, these materials often exhibit a second peak (SP) in field dependent magnetization data apart from a pronounced central peak around $H$ = 0 which is also commonly known as `fishtail' effect. However, the doped Ca-122 materials appear to be quite distinct in this regard where the previous studies have shown SP is surprisingly absent.\cite{ashim-CFCA,civale}

In this work, we have studied the isothermal $M(H)$, $J_c$, flux jumps, fishtail effect and vortex dynamics in a hole-doped 122 Fe-pnictide material i.e., Ba$_{0.65}$Na$_{0.35}$Fe$_2$As$_2$ (BNFA) by means of a study of bulk magnetization with $H||c$ axis. Note, that the corresponding hole-doped sister compound Ba$_{1-x}$K$_x$Fe$_2$As$_2$ is well studied in terms of above mentioned properties. However, even though both compounds are hole-doped, the dopants are different in ionic size and electronic configuration, which might  affect those properties differently. The occurrence of flux jumps is not common in Fe-pnictides, and it has only been reported for 111 and hole-doped 122 materials,\cite{ashim-LiFeAs,wang} yet a detailed study has not been undertaken so far. The material BNFA exhibits reasonably high $T_c$ (29.4 K) and multigap SC as revealed from a recent specific heat study.\cite{ashim-BNFA} We find a pronounced SP in isothermal $M(H)$ and high $J_c$ ($> 10^{6}$ A/cm$^2$ at 2 K) similar to K-doped materials.\cite{yang,sun} At low temperatures, this material exhibits flux jumps which are strongly influenced by changes in temperature and field sweep rate. Moreover, the experimentally observed field for the first flux jump is found significantly higher than the calculated one. This enhanced magnetic stability presumably arises due to strong flux creeping. Our results imply a collective pinning of vortices, and SP in $M(H)$ likely arises due to a crossover in vortex pinning mechanism from an elastic to plastic regime. Based on available data, a vortex dynamics phase diagram has been constructed on temperature-field plane.                  
 
\section{Experimental Details}
Single crystals of Ba$_{0.65}$Na$_{0.35}$Fe$_2$As$_2$ have been prepared using a self-flux method. Details of sample preparation and characterization techniques are described elsewhere.\cite{sai,sai-BFA} It can be mentioned that the crystals used in the present study are the same as in Ref. \onlinecite{ashim-BNFA}. The crystals are characterized with x-ray diffraction (XRD) which confirms the absence of any foreign phases. The chemical composition of the crystal studied in this work has been determined by the energy dispersive x-ray (EDX) analysis spectroscopy conducted at different places of crystal. This study shows that Na is homogeneously distributed within the experimental resolution. The size of the crystal used here is 2.24 $\times$ 1.52 $\times$ 0.14 mm$^3$. The magnetization data have been collected using a SQUID-VSM magnetometer (MPMS, Quantum Design ). The $M(H)$ isotherms are collected after cooling the sample each time in zero magnetic field from far above $T_c$ to the target temperatures. For magnetic relaxation, the sample has been cooled in zero magnetic field to the target temperatures then the magnetic field is applied and magnetization has been measured as a function of time ($t$).      
 
\section{Results and Discussions}
Fig. 1 presents the isothermal $M(H)$ hysteresis loops with $H||c$ axis collected at various temperatures within the superconducting state ($T_c$ = 29.4 K)\cite{ashim-BNFA}. As evident in Fig. 1a, all M(H) isothermals at any temperature exhibit a central peak around $H$ = 0, yet at 2 K the loop shows irregular discontinuities which are known as `flux jumps' and will be discussed later. With increase in field, at low temperatures (2, 5 and 10 K) the width of the loops decreases, however, at higher temperatures it is seen that hysteresis width initially decreases showing a minimum at a field $H_m$ and then again increases (see Fig. 1). This anomaly is more evident in Fig. 1b where the $M(H)$ loops demonstrate another pronounced peak or so called SP at higher field apart from central peak. The SP effect has been observed in superconductors both with low and high transition temperatures and has been studied extensively. Its origin may be attributed to various mechanisms.\cite{roy,daeumling,klein,yoshizaki,kopylov,bugo,elbaum,rosenstein,khaykovich,ertas,giller,radzyner,abulafia} With decreasing temperature, the position of the peak ($H_p$) moves to higher fields, eventually beyond the available field range. This can explain the nonvisibility of SP at low temperatures in Fig. 1a. At high temperatures, irreversibility in magnetization associated with the $M(H)$ loops is seen to vanish above a characteristic field $H_{irr}$. All the $M(H)$ loops are rather symmetric with respect to both field polarity as well as direction of field, implying that bulk pinning plays a dominant role in this compound.            

\begin{figure}
	\centering
		\includegraphics[width=7cm]{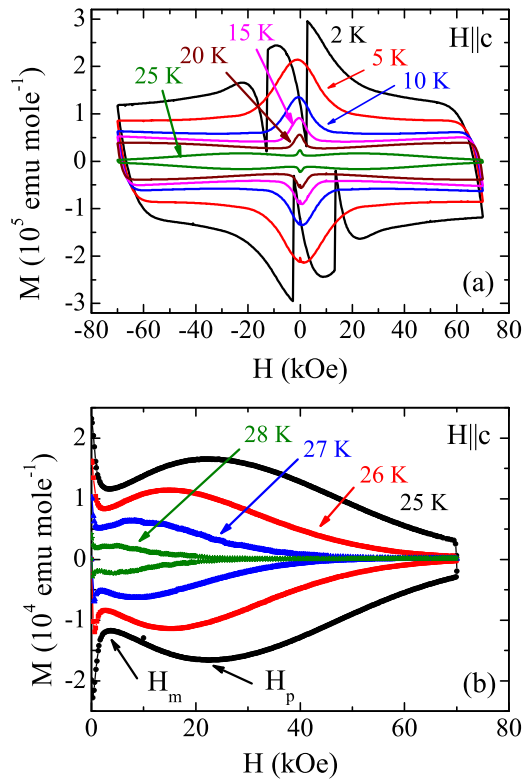}
	\caption{(Color online) Isothermal magnetization hysteresis loops as a function of magnetic field with $H||c$ are shown for Ba$_{0.65}$Na$_{0.35}$Fe$_2$As$_2$ at temperatures (a) 2, 5, 10, 15, 20, 25 K, and (b) 25, 26, 27, 28 K. Plots at high temperatures exhibit a pronounced second peak.}
	\label{fig:Fig1}
\end{figure}

From the magnetic irreversibility in $M(H)$ plots, we have calculated $J_c$ at different temperatures using Bean's critical state model,\cite{bean}

\begin{eqnarray}
	J_c = 20\frac{\Delta M}{a\left(1 - a/3b\right)},
\end{eqnarray}
 
where $\Delta M$ = $M_{dn}$ - $M_{up}$, $M_{dn}$ and $M_{up}$ are the magnetization measured with decreasing and increasing fields, respectively, and $a$ and $b$ ($b > a$) are the dimensions of crystal surface perpendicular to the applied field. The calculated $J_c$ are shown in Fig. 2. As expected from the magnetization plots in Fig. 1, $J_c(H)$ is nonmonotonic at higher temperatures and shows a pronounced SP. With lowering temperature, $J_c$ increases, however, at 2 K it exhibits jumps. Note, that at low temperatures there is only an insignificant change in $J_c(H)$ in the available field range (up to 70 kOe). The determined $J_c$ are rather high and compare well with K-doped compounds i.e., Ba$_{1-x}$K$_x$Fe$_2$As$_2$ where at low temperatures $J_c$ $>$ 10$^6$ A/cm$^2$.\cite{yang,sun}   

\begin{figure}
	\centering
		\includegraphics[width=7cm]{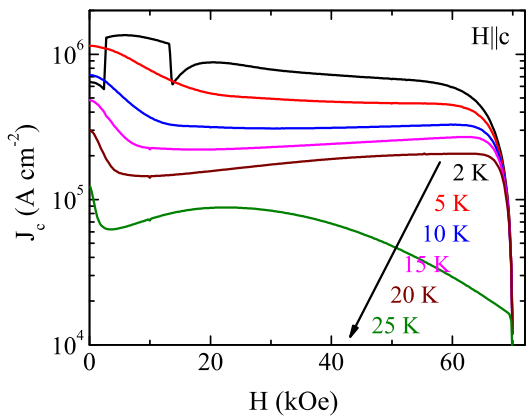}
	\caption{(Color online) Critical current density vs magnetic field with $H||c$ for Ba$_{0.65}$Na$_{0.35}$Fe$_2$As$_2$ at temperatures 2, 5, 10, 15, 20 and 25 K.}
	\label{fig:Fig2}
\end{figure}

To examine the temperature dependence, we have plotted $J_c$ vs $T$ in the main panel of Fig. 3. As evident in Fig. 3, $J_c(T)$ at $H \sim$ 0 decreases monotonically. Apparently, there is a different slope for $J_c(T)$ in low and high temperature regime. For the sake of clarity, we have marked those two regimes by two distinct straight solid lines in Fig. 3. We tried to fit our data with a commonly used function i.e., $J_c(T)$ = $J_c(0)(1 - T/T_c)^n$ which yields $J_c(0)$ = 1.55 $\times$ 10$^6$ A/cm$^2$ and $n$ = 1.73. Unlike Co-doped BaFe$_2$As$_2$,\cite{prozorov} we can not fit our data in the full temperature range with one function (dashed line in Fig. 3). A similar behavior has also been observed for Fe-pnictide 1111 materials PrFeAsO$_{0.9}$ and NdFeAsO$_{0.9}$F$_{0.1}$, where the measured local (high) $J_c$ exhibits a change in temperature dependence.\cite{beek-1111} This behavior has been attributed by authors to an additive effect of weak collective pinning by dopant atoms and strong pinning by a spatial variation in dopant atom density. The strong pinning is active in low field regime and has weaker temperature dependence than the weak collective pinning. The signature of strong pinning is evident in the field dependence of $J_c$. For instance, at low field $J_c(H)$ shows a plateau which is followed by a steep decrease of power-law behavior i.e., $J_c \propto H^{-\beta}$ with $\beta$ = 0.5 $< \beta <$ 0.63.\cite{beek-SP,ovchiv} In order to check the presence of strong pinning also in our sample, we have plotted $J_c$ vs $H$ on log-log scale at different temperatures, and estimated the exponent $\beta$. The inset of Fig. 3 exemplarily shows such a plot for the data at 10 K. With increasing field it is observed that, initially $J_c(H)$ exhibits a plateau, then decreases $\propto$ $H^{-0.54}$ and at higher field it almost saturates at a value $J_c$-sat which is quite similar to strong pinning model described in Refs. \onlinecite{beek-1111,beek-SP,ovchiv}. We have plotted the temperature variation of $J_c$-sat in main panel of Fig. 3 which exhibits linear behavior down to 10 K (at 5 K, $J_c$-sat is not achieved within the available field range). It is worth mentioning that with increase in temperature, $\beta$ decreases where within the temperature span of 5 - 28 K, $\beta$ changes from 0.56 to 0.3. This decline in $\beta$ may be due to an increased flux creep at elevated temperatures. Nonetheless, this observation primarily shows the presence of strong pinning in this compound additionally, while the major contribution to $J_c$ comes from weak collective pinning, as previously observed in doped 1111 and 122 families.\cite{beek-1111,beek-prl}      

\begin{figure}
	\centering
		\includegraphics[width=7cm]{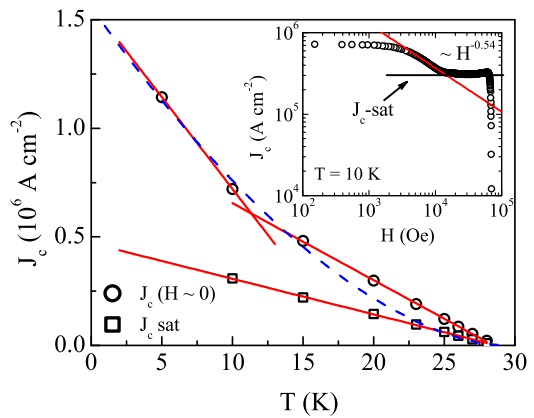}
	\caption{(Color online) Temperature dependence of supercurrent density shown for Ba$_{0.65}$Na$_{0.35}$Fe$_2$As$_2$. Critical current density calculated from Eq. 1 at $H$ $\sim$ 0 is represented by circles and that at high field saturated state (depicted in inset) are represented by squares. Red straight lines are linear fits to the data. The blue dashed line is a fit to the data using the following function $J_c(T)$ = $J_c(0)(1 - T/T_c)^n$. Inset shows double logarithmic plot of critical current density vs field at 10 K. The line is linear fit to data.}
	\label{fig:Fig3}
\end{figure}
 
\begin{figure}[b]
	\centering
		\includegraphics[width=7cm]{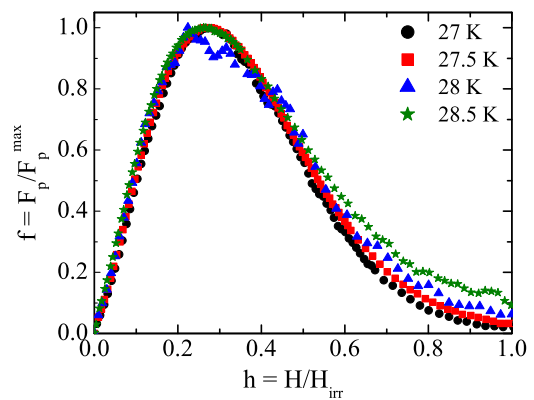}
	\caption{Normalized flux pinning force $f_p = F_p/F_p^{max}$ is shown as a function of reduced field $h = H/H_{irr}$ at temperatures 27, 27.5, 28, 28.5 K for Ba$_{0.65}$Na$_{0.35}$Fe$_2$As$_2$.}
	\label{fig:Fig4}
\end{figure}

Furthermore, to understand the vortex pinning mechanism in this compound we have calculated the pinning force ($F_p$) from the critical current density and applied field, using $F_p$ = $HJ_c$ at different temperatures. It has been shown by Dew-Hughes\cite{dew} that in the scenario of single vortex pinning mechanism the normalized pinning force $f_p$ = $F_p/F_p^{max}$ as a function of reduced field $h$ = $H/H_{c2}$ obeys a scaling relation i.e., $f_p \propto h^p(1 - h)^q$ where $F_p^{max}$ is the maximum pinning force, and $p$ and $q$ are the exponents. This implies that the plotting of $f_p$ vs $h$ at different temperatures will fall on a single curve for a given sample. The position of peak ($h_{max}$) in $f_p(h)$ as well as the extracted fitting parameters $p$ and $q$ provide information about the nature of pinning and its origin. In our analysis we have, however, used $H_{irr}$, the irreversible magnetic field at which $J_c(H)$ $\rightarrow$ 0, instead of $H_{c2}$ as commonly used for cuprates and Fe-pnictides. In Fig. 4 we show the $f_p(h)$ plot for 27, 27.5, 28 and 28.5 K for which $H_{irr}$ values are available to us. A reasonably nice scaling of the data is observed at temperatures close to $T_c$ with $h_{max}$ $\sim$ 0.28. In general, for type-II superconductors the nature of pinning is classified in two categories: $\delta l$ and $\delta T_c$ pinning. The $\delta l$-type pinning arises from a spatial variation in the mean free path of charge carriers while the $\delta T_c$-type pinning is caused by a spatial variation in critical temperature.\cite{blatter} Moreover, for $\delta l$ pinning the defects are small, point sized but in case of $\delta T_c$ pinning the defects are extended, usually larger than the coherence length ($\xi$). According to Dew-Hughes model,\cite{dew} a low value of $h_{max}$ indicates a $\delta l$ type pinning whereas the higher value of $h_{max}$ ($>$ 0.5) corresponds to $\delta T_c$ pinning. Therefore, a low value of $h_{max}$ (0.28) in the studied single crystal of  Ba$_{0.65}$Na$_{0.35}$Fe$_2$As$_2$ suggests that pinning in this material is, primarily, due to the presence of a large density of point like defect centers.\cite{kramer} It is not clear the exact source of defects with the present dataset, however, the dopant atoms and/or the As deficiency could be the possible candidates. At the same time, it is worthy to mention that $h_{max}$ remains highly sensitive to a proper determination of $H_{irr}$. In the case of Fe-pnictides, a wide variation in $h_{max}$ is observed. For instance, $h_{max}$ is found to be 0.33 and 0.43 for hole-doped 122-compounds and 0.37 and 0.45 for electron-doped ones where the disorder mostly has been related to the compositional inhomogeneities.\cite{yang,yama,sun}

\begin{figure}
	\centering
		\includegraphics[width=7cm]{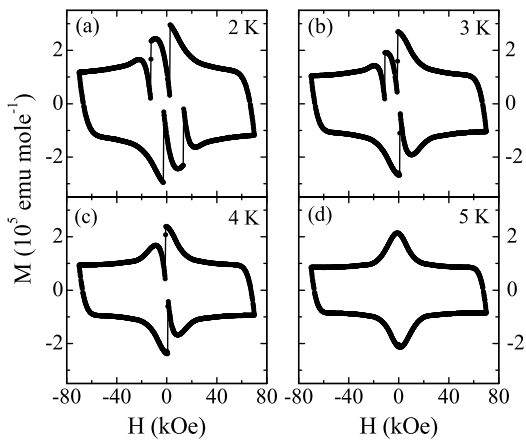}
	\caption{Field dependence of magnetic hysteresis loops with field sweep rate 80 Oe/s for Ba$_{0.65}$Na$_{0.35}$Fe$_2$As$_2$ at (a) 2 K, (b) 3 K, (c) 4 K, and (d) 5 K.}
	\label{fig:Fig5}
\end{figure}
 
To understand the anomalous flux jumps at low temperature (Fig. 1) we have collected $M(H)$ data at narrow temperature interval of 2 - 5 K with $H||c$ axis. Fig. 5 presents the envelope curves of $M(H)$ (70 to -70 kOe and then back to 70 kOe) at 2, 3, 4, and 5 K with field sweep rate of 80 Oe/s. Flux jumps are evident only up to 4 K. Note, that the jumps are not a fully reproducible in terms of both their occurrence and positions. Also, the jumps are only observed in the increasing field branch of both field directions with this sweep rate, and with increasing temperature the number of jumps reduces (see Fig. 5). It is also noticeable that after each jump magnetization approaches to zero, implying that jumps are almost complete and the material is driven close to the normal state.   
  
The sweep rate of the applied magnetic field has been found to have a profound effect on the flux jump phenomenon. Fig. 6 depicts the measured $M(H)$ envelope curves at 2 K with different sweep rates i.e., 15, 30, 80, 200, 300 and 400 Oe/s. At low sweep rates the jumps are more complete than at higher sweeping rates. With increasing sweep rate up to 200 Oe/s, the number of jumps remains the same, though their positions change, but the jumps become more numerous at a higher sweep rate. Moreover, in contrast to low sweep rate measurements where the jumps are only observed in increasing field branch, the higher sweep rate data depict jumps in both increasing as well as decreasing branches of applied field. In Fig. 7, we present the virgin curves corresponding to envelope curves shown in Fig. 6. Though the jumps do not follow a regular pattern with field sweep rates but they disappear at low sweep rate which has also been observed in cuprates and other superconductors.\cite{henry,nabialek,tien}
 
\begin{figure}
	\centering
		\includegraphics[width=7cm]{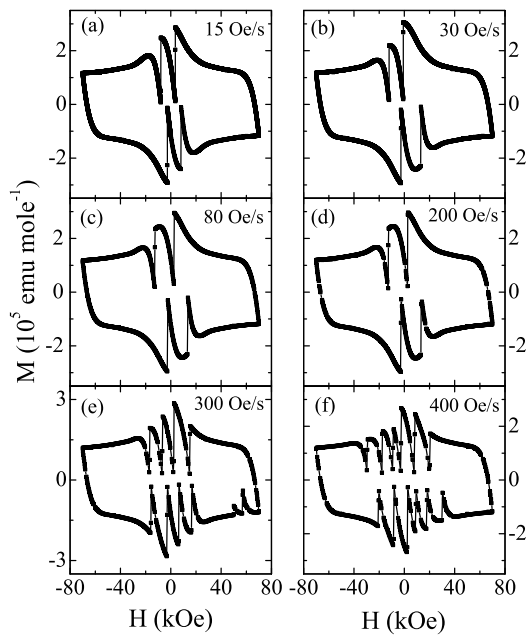}
	\caption{Field dependence of magnetic hysteresis loops (envelope curves) at 2 K for Ba$_{0.65}$Na$_{0.35}$Fe$_2$As$_2$ with field sweep rate of (a) 15 Oe/s (b) 30 Oe/s (c) 80 Oe/s (d) 200 Oe/s (e) 300 Oe/s (f) 400 Oe/s.}
	\label{fig:Fig6}
\end{figure}

On the theoretical background, flux jumps or magnetic instability was explained long ago within the framework of adiabatic model.\cite{swartz} This model argues that a flux jump instability occurs when the dissipative heat generated by movement of fluxes is not able to be fully absorbed by the sample, thus raising its temperature. To be precise, this local adiabatic condition is fulfilled if the thermal diffusivity ($D_t$) of the sample is lower than the magnetic diffusivity ($D_m$). In this situation, a small thermal fluctuation arising from dissipation causes a reduction in the screening current which allows a penetration of more flux lines in the sample. This movement of fluxes in turn produces dissipation heat, hence further heating the sample. This cyclic feedback forms an avalanche-like process leading to flux jumps. The adiabatic model further predicts the instability field for the first flux jump $H_{fj}$ as,

\begin{figure}
	\centering
		\includegraphics[width=7cm]{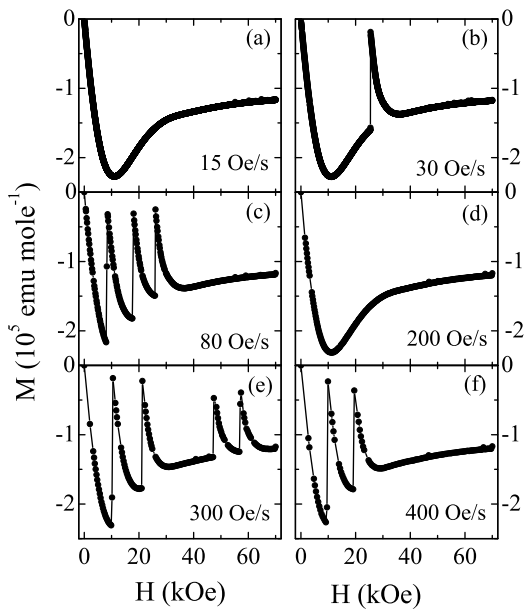}
	\caption{Virgin magnetization curves vs magnetic field collected at 2 K are shown for Ba$_{0.65}$Na$_{0.35}$Fe$_2$As$_2$ with field sweep rate of (a) 15 Oe/s (b) 30 Oe/s (c) 80 Oe/s (d) 200 Oe/s (e) 300 Oe/s (f) 400 Oe/s.}
	\label{fig:Fig7}
\end{figure}

\begin{eqnarray}
	H_{fj} = \sqrt{\frac{2\mu_0CJ_c}{-dJ_c/dT}}
\end{eqnarray}
       
where $\mu_0$ is the magnetic permeability of vacuum and $C$ is the specific heat. Taking $C$ of this compound at 2 K to be 152.81 J K$^{-1}$ m$^{-3}$ (Ref. \onlinecite{ashim-BNFA}) and $dJ_c/dT$ from Fig. 3, we calculate $H_{fj}$ to be about 768 Oe. In Fig. 8 we have plotted the experimentally observed $H_{fj}$ as collected from Fig. 7 against the field sweep rate. It follows from Fig. 8 that at higher sweep rates the $H_{fj}$ does not vary substantially, whereas it increases significantly at a lower sweep rates vanishing at 15 Oe/s. Interestingly, the observed $H_{fj}$ in Fig. 8 is more than one order of magnitude higher than the calculated one mentioned above. This implies that material is quite stable against the flux jumps. It can be mentioned that such disagreement between the calculated $H_{fj}$ following adiabatic model and the experimentally observed $H_{fj}$ has earlier been shown for cuprate superconductors such as La$_{1.86}$Sr$_{0.14}$CuO$_4$ (Ref. \onlinecite{henry}) and Bi$_2$Sr$_2$CaCu$_2$O$_{8+\delta}$ (Ref. \onlinecite{nabialek}). However, the dependence of $H_{fj}$ on the magnetic field sweep rate in Fig. 8 is an indicative of flux instability being related to the vortex creep phenomenon. The creeping of flux is likely to modify the field profile in the superconductors while sweeping the magnetic field. It is, indeed, shown by McHenry\cite{henry} that heat generated due to flux motion in absence of flux creep is higher than that in presence of flux creeping. To look into the flux creep event in this material, we have measured the magnetic relaxation. For that the sample has been cooled in zero magnetic field to 2 K, and after proper thermal stabilization a field of 10 kOe ($H||c$) is applied, and magnetization has been measured as a function of time. The inset of Fig. 8 presents the measured $M(t)$ data which exhibits a sizable increase in magnetization ($\sim$ 7.4\%) over a period of 2 h. We calculate a fairly high normalized relaxation rate $S$ (d$\ln M$/d$\ln t$) = 0.013 at this low temperature. This implies that flux creep is quite significant in this material and may be the cause for enhanced stability in flux flow.       
 
\begin{figure}
	\centering
		\includegraphics[width=7cm]{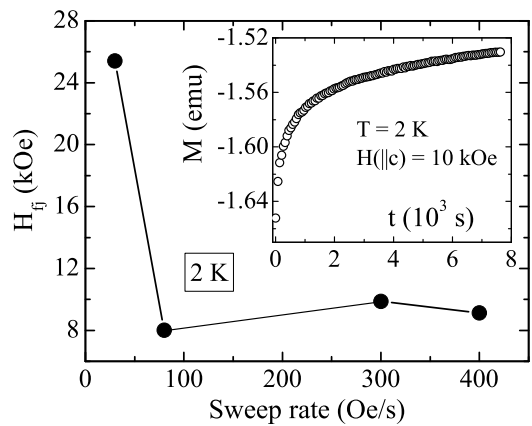}
	\caption{The dependence of first jump field $H_{fj}$ on the field sweep rate is shown at 2 K for Ba$_{0.65}$Na$_{0.35}$Fe$_2$As$_2$. The inset shows magnetic relaxation at 2 K in field $H||c$ = 10 kOe.}
	\label{fig:Fig8}
\end{figure}

As mentioned before and depicted in Fig. 1, this material exhibits a SP in $M(H)$. To understand the origin of SP we have performed detailed magnetic relaxation measurements at constant temperature with different applied magnetic fields.\cite{yeshu,blatter} The origin of magnetic relaxation in superconductors in presence of field is a nonequilibrium spatial distribution of vortices due to pinning. The Lorentz force experienced by the vortices coupled with thermal fluctuations drives the vortices in to the bulk of superconductors, hence causing an evolution in $M(t)$. Following a nonlinear dependence of pinning potential $U$ upon the current density $J$ i.e., $U(J) = \left(U_0/\mu\right)[(J_c/J)^\mu - 1]$, where $U_0$ is the barrier potential in absence of driving force and $\mu$ is the field and temperature dependent exponent, the magnetic relaxation (taking $M$ is proportional to $J$) can be explained using an interpolation formula,\cite{yeshu}

\begin{eqnarray}
	M(t) = M_0\left[1+ \frac{\mu k_BT}{U_0}\ln\left(\frac{t}{t_0}\right)\right]^{-1/\mu}
\end{eqnarray}

Here $M_0$ is the initial magnetization, $k_B$ is the Boltzmann constant and $t_0$ is the macroscopic characteristic time which depends on the sample size and shape which is usually much shorter than the observation time.\cite{feigel} From Eq. 3, the functional form of $S(t)$ comes out as,

\begin{eqnarray}
	S(t) = \frac{k_BT}{U_0 + \mu k_B T \ln\left(t/t_0\right)}
\end{eqnarray}

\begin{figure}
	\centering
		\includegraphics[width=8cm]{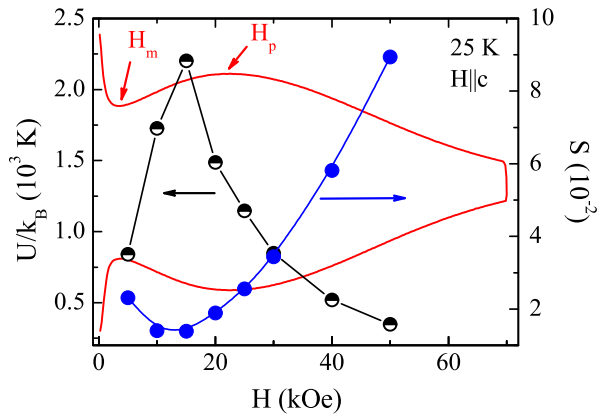}
	\caption{Field dependence of pinning potential $U_0/k_B$ and normalized relaxation rate $S$ ($t$ = 1000 s) for Ba$_{0.65}$Na$_{0.35}$Fe$_2$As$_2$ at 25 K. The continuous line represents the $M(H)$ plot at 25 K where data have been plotted without a vertical scale.}
	\label{fig:Fig9}
\end{figure}

We have measured $M(t)$ in different magnetic fields for about 8000 s at 25 K where the second peak is clearly evident within the available field range (see Fig. 1). By fitting Eq. 3 with our experimental data $M(t)$, we have extracted parameters such as, $M_0$, $\mu$, $U_0/k_B$ and $t_0$ at each field, and using Eq. 4, we have calculated $S(t)$ at $t$ = 1000 s as a representative example. Fig. 9 shows the field dependence of $U_0/k_B$ and $S$ along with the magnetization plot. As evident, with increasing field the $U_0/k_B$ initially increases and then decreases showing a peak around 15 kOe. The $S(H)$, however, exhibits an opposite trend of $U_0/k_B$ which is evident from Eq. 4. The value of $S$ for the present compound is higher than for conventional superconductors, however, comparable with cuprates and other Fe-pnictides.\cite{yeshu,prozorov,ashim-CFCA,ashim-LiFeAs,kope} Often, the appearance of SP in $M(H)$ has been ascribed to a crossover in vortex creep mechanism from elastic to plastic creep with field.\cite{abulafia,giller,prozorov} In this model, the pinning potential $U$ has been shown to have different dependence on $H$.\cite{abulafia} While in case of elastic creep the barrier potential follows $U \propto H^\nu$ where $\nu$ is a positive number, increasing with field. On the other hand, for plastic creep it shows $U \propto 1/\sqrt{H}$, decreasing with field. The creeping is governed by the mechanism that has a lower potential, which straightforwardly implies that elastic and plastic creep is active in low and high fields, respectively. Also, the position of SP is connected to the matching of these two potentials. The behavior $U(H)/k_B$ in Fig. 9 where it increases and then decreases with a peak close to SP in $M(H)$ is quite suggestive that origin of SP in this material is most likely related to crossover in creeping mechanism discussed above. However, there have been other models proposed to explain the SP effect like, structural phase transition in vortex lattice,\cite{rosenstein} and this has recently been shown for Fe-pnictide material Ba(Fe$_{0.925}$Co$_{0.075}$)$_2$As$_2$.\cite{kope} Therefore, for a deeper understanding of the origin of SP in this new superconducting materials, further investigations employing microscopic measurement tools such as, neutron scattering, scanning tunneling microscope, etc. are necessary.     

\begin{figure}
	\centering
		\includegraphics[width=7cm]{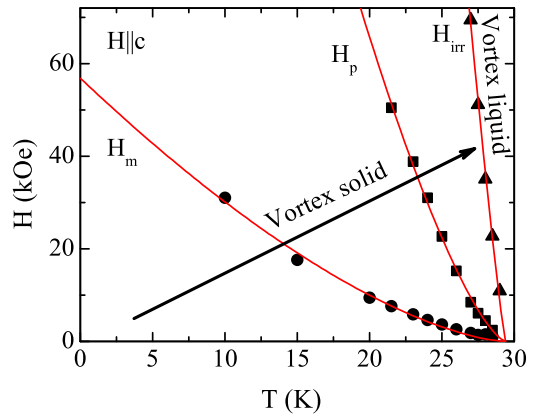}
	\caption{Vortex phase diagram on field-temperature plane is shown for Ba$_{0.65}$Na$_{0.35}$Fe$_2$As$_2$ with $H||c$ axis. The solid lines represents the fit to the experimental data with functional form $H_x(T) = H_x(0)(1 - T/T_c)^n$ (see text).}
	\label{fig:Fig10}
\end{figure}
 
Finally, with the above discussed experimental results we have constructed a vortex phase diagram on the field-temperature plane (Fig. 10). We have mainly plotted the temperature dependence of three characteristic field i.e., $H_{irr}$, $H_p$ and $H_m$. Above the irreversible field $H_{irr}$, vortices are in an unpinned liquid state. Below the irreversibility line a solid vortex lattice is realized though its nature modifies as a function of temperature and field. The SP field $H_p$ presumably marks here the crossover from elastic to plastic creep regime and $H_m$ marks the onset of SP. It is clear in Fig. 10 that both $H_m$ and $H_p$ exhibit a strong temperature dependence, however, the $H_{irr}(T)$ is quite steep implying that a large part of vortex phase diagram is interesting from the viewpoint of applications. The temperature dependence of all the characteristic fields can be well explained by functional form, $H_x(T) = H_x(0)(1 - T/T_c)^n$ where $n$ is an exponent. Upon fitting the different set of experimental data in Fig. 10, we obtain $H_m(0)$ = 5.7 T and $n$ = 1.52, $H_p(0)$ = 33.1 T and $n$ = 1.42, $H_{irr}(0)$ = 120.6 T and $n$ = 1.15. The obtained values of exponent $n$ in present study are quite similar to other Fe-pnictide materials.\cite{prozorov,shen,ashim-LiFeAs,yang} It is also noticeable that extracted $H_p(0)$ and $H_{irr}(0)$ in hole doped 122 Fe-pnictides in general are significantly higher than in electron doped materials,\cite{prozorov,yang} which is justified by their higher $H_{c2}$ ($\sim$ 150 T) compared to electron doped materials.\cite{welp,yama}            
 
\section{Conclusions}
In summary, we have measured the isothermal magnetization loops $M(H)$ and the magnetic relaxation curves $M(t)$ in single crystalline samples of the hole-doped Fe-pnictide high temperature superconductor Ba$_{0.65}$Na$_{0.35}$Fe$_2$As$_2$. The $M(H)$ loops exhibit a so-called fishtail anomaly (or a second peak - SP), which is pronounced up to temperatures close to $T_c$. In addition, remarkable flux jumps are observed in $M(H)$ at low temperatures. The critical current densities $J_c$ calculated form the $M(H)$ dependences  are reasonably high and compare well with other hole-doped 122 compounds. The analysis of temperature- and field-dependent $J_c$ indicates that collective (weak) pinning of vortices by dense microscopic point defects is mainly responsible for high values of $J_c$, though a strong pinning mechanism is likely also to give some contribution. The field sweep rate has strong influence on the observed flux jumps. By lowering the sweep rate the number of jumps reduces and the field for the first jump $H_{fj}$ increases. However, $H_{fj}$ calculated following the adiabatic model shows a much lower value than the experimentally observed one, which is probably due to an effect of flux creeping. The results of field dependent magnetic relaxation measurements support applicability of the collective pinning model to the studied compound, and further suggest that SP in $M(H)$ probably arises due to crossover from elastic to plastic flux creep mechanism with field. Based on the available data, we have constructed a vortex phase diagram on the field-temperature plane which highlights that the vortex liquid phase in this material is confined to a small region of the phase diagram only. The major part of the vortex diagram is dominated by the vortex solid phase, a feature highly appreciative for practical applications.
 
\section{Acknowledgment}
AKP is thankful to Cornelis J. van der Beek and Marcin Konczykowski for fruitful discussions. SW acknowledges support by DFG under the Emmy-Noether program (Grant No. WU595/3-1). We thank M. Deutschmann, S. M\"uller-Litvanyi, R. M\"uller, J. Werner, and S.~Ga{\ss} for technical support. This work has been supported by the DFG, Germany through grant no BE 1749/13 and WO 1532/3-1.

\end{document}